\documentclass[aps,pra,superscriptaddress,showpacs,twocolumn,10pt]{revtex4-1}
\usepackage{bm}
\usepackage{amsmath}
\usepackage{textcomp}
\allowdisplaybreaks
\usepackage{longtable}
\usepackage{dcolumn}
\newcolumntype{.}{D{x}{}{-1}}

\newcommand*{\centt}[1]{\multicolumn{1}{c}{#1}}
\newcolumntype{w}[1]{D{.}{.}{#1}}

\newcommand{\lbr}{\langle}
\newcommand{\rbr}{\rangle}

\usepackage{graphicx}
\usepackage{xcolor}

\begin{document}

\title{QED calculation of ionization energies of $\bm{1snd}$ states in helium}

\author{Vladimir A. Yerokhin}
\affiliation{Center for Advanced Studies, Peter the Great St.~Petersburg Polytechnic University,
Polytekhnicheskaya 29, 195251 St.~Petersburg, Russia}

\author{Vojt\v{e}ch Patk\'o\v{s}}
\affiliation{Faculty of Mathematics and Physics, Charles University,  Ke Karlovu 3, 121 16 Prague
2, Czech Republic}

\author{Mariusz Puchalski}
\affiliation{Faculty of Chemistry, Adam Mickiewicz University, Umultowska 89b, 61-614 Pozna{\'n},
Poland}

\author{Krzysztof Pachucki}
\affiliation{Faculty of Physics, University of Warsaw,
             Pasteura 5, 02-093 Warsaw, Poland}

\begin{abstract}

Quantum electrodynamical (QED) calculations of ionization energies of the $1snd\,D$ states are
performed for the helium atom. We reproduce the previously known relativistic and QED effects
up to order $m\alpha^5$ and extend the theory by calculating the complete $m\alpha^6$
correction. The total contribution of the $m\alpha^6$ effects is shown to be much smaller than
previously estimated, due to a large cancelation between the radiative and non-radiative parts
of this correction. As a result of our calculations, we confirm the previously reported
deviations between measured transition energies and theoretical predictions for the $nD$--$2S$
and $nD$--$2P$ transitions. Possible reasons for this discrepancy are analyzed.

\end{abstract}

\maketitle

\section{Introduction}

Helium is the simplest few-electron atom, and it has played a special role in physics since the early days
of quantum mechanics. The relativistic effects in helium are smaller than the nonrelativistic
energy roughly by a factor of $(Z\alpha)^2$ (where $Z=2$ is the helium nuclear charge number and $\alpha$ is the
fine-structure constant) and thus can be accounted for by perturbation theory, yielding a
convenient approach for an accurate theoretical description of helium spectra. Being a monatomic
gas, helium is easy to work with and can be studied experimentally to great accuracy due to
the presence of narrow spectral lines in the spectrum. A large body of experimental results
exists for helium spectra, many of them reaching the relative precision of a few parts in
$10^{-12}$
\cite{zheng:02,rooij:11,luo:13,luo:13:b,notermans:14,luo:16,zheng:17,rengelink:18,kato:18,huang:18}.

Motivated by the experimental success, numerous theoretical calculations have been performed for
transition energies of the helium atom. In particular, extensive calculations by Drake {\em et al.}
\cite{drake:92,drake:98:cjp,morton:06:cjp} covered all states with angular
momentum up to $L = 8$ and principal quantum number up to $n=10$. These calculations
accounted for all QED effects up to order $m\alpha^5$ and partly included some higher-order QED
effects. The first complete calculation of the next-order $m\alpha^6$ effects was accomplished by
one of us for the $n = 1$ and $n = 2$ states of helium
\cite{pachucki:02:jpb,pachucki:06:hesinglet,pachucki:06:he} and later extended to light
helium-like ions \cite{yerokhin:10:helike}.

The present status of the theory of the $n = 1$ and $n = 2$ states of helium was summarized in
a recent review \cite{pachucki:17:heSummary}. The comparison with available experimental data
presented in that work showed good agreement for the intrashell $n=2$ transitions. However,
recent experiments \cite{luo:13:b,luo:16,huang:18} have reported deviations from theoretical
predictions for transitions involving $D$ states. In order to clarify this,
we have performed  in Ref. \cite{wienczek:19} an independent calculation of the helium $3D$
ionization energies, including the complete $m\alpha^6$ correction. As a result, we confirmed
the systematic deviation between the theoretical and experimental $3D$--$2S$,$2P$ transition
energies on the level of $1.5\,$--$\,3$ standard deviations.

In the present work we aim to gain further insight into the reported discrepancy between theory
and experiment, by calculating the ionization energies of the higher excited $nD$ states of
helium, for which theoretical predictions can be much more accurate in terms of the absolute uncertainty.
We will reproduce the previously known QED effects up to order $m\alpha^5$ and extend the
previous theory by calculating the complete $m\alpha^6$ correction.

\section{Calculation}

In the general spirit of the Nonrelativistic QED approach, the energies of the helium atom are
represented as a double expansion in small parameters $\alpha$ and $m/M$, where $\alpha$ is the
fine-structure constant and $m/M$ is the electron-to-nucleus mass ratio. Specifically, we express
the energy $E$ of a $1snd\,D$ state as
\begin{align}
E =   &\ m\Bigg\{\alpha^2 \Big[ E^{(2,0)} + \frac{m}{M}\,E^{(2,1)} + \frac{m^2}{M^2}\,E^{(2,2+)}\Big]
 \nonumber \\
    + &\ \alpha^4 \Big[ E^{(4,0)} + \frac{m}{M}\,E^{(4,1)}\Big]
    +  \alpha^5 \Big[ E^{(5,0)} + \frac{m}{M}\,E^{(5,1)}\Big]
 \nonumber \\
    + &\ \alpha^6  E^{(6,0)} + \alpha^7  E^{(7,0)} + E_{\rm MIX} + E_{\rm FNS}\Bigg\}\,,
\end{align}
where the superscripts $i$ and $j$ in $E^{(i,j)}$ specify the order in $\alpha$ and $m/M$,
respectively.

The individual contributions to energies of the $D$ states were examined in detail in our previous
investigation \cite{wienczek:19}. The leading contribution $E^{(2,0)}$ is the nonrelativistic
energy for the infinitely heavy nucleus. $E^{(2,1)}$ and $E^{(2,2+)}$ are the first-order and
higher-order finite-nuclear-mass corrections to the nonrelativistic energy. The latter contains
nuclear mass contributions $\propto (m/M)^2$ and higher. $E^{(4,0)}$ and $E^{(4,1)}$ are the
leading relativistic (Breit) correction in the nonrecoil limit and the corresponding first-order
recoil correction, respectively.  $E^{(5,0)}$ and $E^{(5,1)}$ are the leading QED correction in the nonrecoil
limit and the corresponding first-order recoil correction, respectively. $E^{(6,0)}$ and $E^{(7,0)}$ are the
higher-order QED corrections. $E_{\rm MIX}$ is the correction due to mixing between the triplet
$n\,^3\!D_2$ and singlet $n\,^1\!D_2$ states. This contribution is enhanced due to a small energy
difference of these states and thus requires a separate treatment, see Ref.~\cite{wienczek:19}
for details. Finally, $E_{\rm FNS}$ is the correction induced by the finite nuclear size.

The main part of the present investigation is the complete calculation of the QED effects of
order $m\alpha^6$, i.e. $E^{(6,0)}$. It is convenient to split this correction into the
radiative and non-radiative parts,
\begin{equation}
  E^{(6,0)} = E^{(6,0)}_{\rm rad} + E^{(6,0)}_{\rm nrad}\,.
\end{equation}
The radiative part is the sum of the one-loop and two-loop radiative contributions,
$E^{(6,0)}_{\rm rad} = E_{R1}+E_{R2}$,
\begin{align}
  E_{R1} &\ = Z^2\,\biggl[\frac{427}{96}-2\,\ln2\biggr]\,
  2\pi\,\bigl<\delta^3(r_1)\bigr>
  \nonumber \\ &
  +\biggl[
  \frac{6\,\zeta(3)}{\pi^2}-\frac{697}{27\,\pi^2}-8\,\ln2+\frac{1099}{72}
  \biggr]\,\pi\,\big<\delta^3(r)\big>,
\end{align}
\begin{align}
  E_{R2} &\ =
  Z\,\biggl[-\frac{9\,\zeta(3)}{4\,\pi^2}-\frac{2179}{648\,\pi^2}+
  \frac{3\,\ln2}{2}-\frac{10}{27}\biggr]\,
  2\pi\,\bigl<\delta^3(r_1)\bigr>
  \nonumber \\ &
    +\biggl[\frac{15\,\zeta(3)}{2\,\pi^2}+\frac{631}{54\,\pi^2}
  -5\,\ln2+\frac{29}{27} \biggr]\,\pi\,\big<\delta^3(r)\big>\,,
\end{align}
which were obtained from Ref. \cite{korobov:01:prl}.
The nonradiative contribution is the remaining part of the $m\alpha^6$ correction. The
corresponding formulas are long and cumbersome; they are presented in Ref.~\cite{wienczek:19} and
will not be repeated here.

The dominant contribution to the radiative correction $E^{(6,0)}_{\rm rad}$ comes from the part
proportional to $\lbr\delta^3(r_1)\rbr$. This part has been known for a long time and was included
in previous calculations, in particular those in Ref.~\cite{morton:06:cjp}. Conversely, the
nonradiative $m\alpha^6$ correction was not known until recently and thus defined the
uncertainty of theoretical predictions in the helium atom and in light helium-like ions.

The first complete calculation of the $m\alpha^6$ effects was accomplished for the $n = 1$ and $n
= 2$ states of helium by one of us in
Refs.~\cite{pachucki:02:jpb,pachucki:06:hesinglet,pachucki:06:he}. This calculation was later
extended to light helium-like ions \cite{yerokhin:10:helike} and to the hydrogen molecule
\cite{puchalski:16}. In our previous investigation we performed a calculation of the $m\alpha^6$
effects for the helium $3D$ states \cite{wienczek:19}, and the main part of the present investigation
is to extend this calculation to the higher excited $nD$ states with $n = 4$, 5, and 6.

The numerical approach of the evaluation is described in detail in Ref.~\cite{wienczek:19}. In
the present work we employed increased computer resources in order to improve the numerical
accuracy for the $n = 3$ state and to ensure the controllable convergence of the results for the
higher excited states. Numerical results of our calculations of the $m\alpha^6$ corrections for
ionization energies of $1snd$ states of $^4$He are presented in Table~\ref{tab:ma6}.

We find that both the radiative and nonradiative $m\alpha^6$ corrections to the ionization energy
for higher excited $n$ states decrease as $1/n^3$, which is the typical scaling for most of the
QED effects. Much less typical is the fact that the radiative and nonradiative contributions for
the $D$ states are of the same order of magnitude. In most calculations performed so far, the
radiative effects were found to yield the dominant contribution, whereas the nonradiative QED
effects originating from the electron-electron interaction were rather small. In particular,
Morton {\em et al.}~\cite{morton:06:cjp} estimated the omitted nonradiative $m\alpha^6$ effects
as 10\% of the $m\alpha^6$ radiative contribution. Our numerical calculations summarized in
Table~\ref{tab:ma6} show that the typical pattern of magnitudes of the radiative versus
nonradiative effects is broken for the helium $D$ states, the nonradiative part being as large as
the radiative one and of the opposite sign.

Having calculated the $m\alpha^6$ effects for a series of $nD$ states, we are able to analyse the
$n$ dependence of the corresponding corrections. Specifically, we fit the numerical results
listed in Table~\ref{tab:ma6} to the $1/n$ asymptotic expansion of the following form
\begin{align}\label{eq:ass}
 E^{(6)} = \frac{1}{n^3} \sum_{i = 0}^2 \frac{c_i}{n^i}\,.
\end{align}
The obtained coefficients $c_i$ are presented in Table~\ref{tab:ass}. Using these results, we can
compute the $m\alpha^6$ correction for the $nD$ states with $n>6$.

Table~\ref{tab:energy} summarizes our calculations of the ionization energies of the $4D$, $5D$,
and $6D$ states in helium. Results include QED effects of order $m\alpha^5$ and
$m\alpha^6$. In particular, we performed computations of the Bethe logarithms for the $4D$, $5D$,
and $6D$ states, and our results agree with previous Bethe-logarithm calculations
\cite{drake:01,korobov:19:bethelog}. Our calculations of the ionization energies are complete up
to order $m\alpha^6$, while the QED effects of order $m\alpha^7$ were estimated by scaling the
hydrogenic results for the one-, two-, and three-loop QED effects, as described in
Ref.~\cite{wienczek:19}. The uncertainty of these estimations was taken as 100\%.

The comparison of our total results with the previous theoretical predictions by Morton {\em et
al.}~\cite{morton:06:cjp} demonstrates that the difference comes exclusively from the
non-radiative $m\alpha^6$ QED effects not included in the previous calculation. Apart from this
addition, the consistency between two independent calculations is nearly perfect, the residual
differences being on the level of just 1-3~kHz. However, as already mentioned, the magnitude of
the nonradiative $m\alpha^6$ effects calculated in this work turned out to be much larger than
previously expected, which resulted in a shift of the theoretical values of
Ref.~\cite{morton:06:cjp} by about 10 times their estimated uncertainty.

In the present work we performed explicit numerical calculations for the $nD$ states with $n$ up
to $n = 6$. With the help of the approximate formula (\ref{eq:ass}), we can obtain the
$m\alpha^6$ effects also for the higher-$n$ states. In view of the great consistency between our
present calculation and the one by Morton {\em et al.}~\cite{morton:06:cjp}, we update their
results for the $nD$ states with $n=7$-10, by adding the nonradiative $m\alpha^6$ correction as
obtained from Eq.~(\ref{eq:ass}) and rescaling their results with the latest value of the Rydberg
constant \cite{NIST}.

The list of the final theoretical ionization energies of $nD$ states of $^4$He for $n\le 10$ is
presented in Table~\ref{tab:energylist}. The small difference of the $3D$ energies from those
reported in our previous work \cite{wienczek:19} is due to the updated value of the Rydberg
constant. The actual Rydberg constant \cite{NIST} differs from the previous CODATA 2014 value
\cite{mohr:16:codata} by about 5 standard deviations, as a result of recent measurements of the
Lamb shift in hydrogen \cite{beyer:17,bezginov:19} and the electron-proton scattering
\cite{xiong:19}. The change of the Rydberg constant induced a shift of the theoretical $3D$
ionization energies of about 10~kHz, about half of the theoretical uncertainty.

Table~\ref{tab:comp} compares the present theoretical values with the available experimental
results for various transitions involving $nD$ states. We observe that the theoretical
predictions for the $nD$--$2S$ and $nD$--$2P$ transitions are systematically larger than the
measured values, the difference varying from 1.5 to 3\,$\sigma$. This is in contrast to the
situation for the intrashell $n = 2$ transitions, where theoretical predictions are in good
agreement with experimental observations \cite{pachucki:17:heSummary}.

Analyzing possible reasons for the deviations between theory and experiment summarized in
Table~\ref{tab:comp}, we first observe that the theoretical energies of the $nD$ states should be
considered as well established on the 1~MHz level. Indeed, the nonradiative $m\alpha^6$ effects
calculated only in this work and not confirmed independently contribute only $0.3$~MHz for the
$3D$ states and much less (decreasing as $1/n^3$) for the higher-$n$ $D$ states. All other
theoretical contributions for the $nD$ states are cross-checked by two independent calculations
and agree up to a few kHz. Therefore, we must assume that the reason for the deviations comes
from the theoretical energies of the $n = 2$ states. Bearing in mind the good agreement between
theory and experiment for the intrashell $n = 2$ transitions \cite{pachucki:17:heSummary}, we
might expect an unaccounted-for contribution of the order of 1~MHz that is nearly the same for
the $S$ and $P$ states and, even more unusually, for the triplet and singlet states.

Theoretical energies of the $n = 2$ states are independently checked up to order $m\alpha^5$.
Therefore, possible unaccounted-for contributions could come either from the $m\alpha^6$ effects
(whose calculations \cite{pachucki:06:hesinglet,pachucki:06:he} have not been confirmed
independently yet) or from the unknown $m\alpha^7$ effects. The latter possibility is likely to
be tested soon, when the project of calculations of all $m\alpha^7$ QED effects for the triplet
states of helium is completed \cite{yerokhin:18:betherel,patkos:20}.

\section{Summary}

We performed calculations of relativistic and QED effects to the ionization energies
of $1snd\,D$ states in helium. Consequently, we reproduced the previously known
relativistic and QED effects up to order $m\alpha^5$ and extended the previous theory by
evaluating the complete $m\alpha^6$ correction. We found that the radiative and nonradiative
$m\alpha^6$ contributions for the $1snd\,D$ states are of the same magnitude and of different
sign, thus cancelling each other to a large extent. This is very different from the
situation for the $S$ and $P$ states, where the radiative part is by far the dominant one.

As a consequence of our calculations, we confirm the previously reported deviations between
measured transition energies and theoretical predictions for the $nD$--$2S$ and $nD$--$2P$
transitions on the level of 1.5--3 standard deviations. The reason for this disagreement is not
known. The most plausible explanation would involve a missing contribution in the theoretical
prediction for the $n = 2$ states, a part of which is not cross-checked by independent
calculations (the $m\alpha^6$ effects) or not yet calculated (the $m\alpha^7$ effects). We
conclude that further theoretical investigations of higher-order QED effects in helium are needed
in order to find out the exact reason for the observed discrepancy.

\section*{Acknowledgement}
V.A.Y. acknowledges support from the Russian Science Foundation (Grant No. 20-62-46006). The work
of K.P., M.P., and V.P  was supported from the National Science Center (Poland) Grant No.
2017/27/B/ST2/02459. Additionally, V.P. acknowledges support from the Czech Science Foundation -
GA\v{C}R (Grant No. P209/18-00918S)

\begin{table*}
\caption{Radiative (rad) and non-radiative (nrad) $m\alpha^6$ corrections for ionization energies
of $1snd$ states of $^4$He, in units of $10^{-3}\,m\alpha^6$.
Conversion factor to MHz is $0.018658054$. \label{tab:ma6}}
\begin{ruledtabular}
\begin{tabular}{lldddd}
$n$ &Term &  \multicolumn{1}{c}{$n\,{}^1\!D$}
                          & \multicolumn{1}{c}{$n\,{}^3\!D_1$}
                                                   & \multicolumn{1}{c}{$n\,{}^3\!D_2$} &
                                                                    \multicolumn{1}{c}{$n\,{}^3\!D_3$} \\
\hline\\[-5pt]
3 & rad   &  -10.7644             & -13.3384             & -13.3384             & -13.3384             \\
  & nrad  &   18.3532\,(8)        &  14.2110\,(40)       &  19.1548\,(12)       &  22.5100\,(42)       \\
  &  sum  &    7.5888\,(8)        &   0.8727\,(40)       &   5.8164\,(12)       &   9.1717\,(42)       \\[3pt]
4 & rad   &   -4.9107             &  -6.3733             &  -6.3733             &  -6.3733             \\
  & nrad  &    7.6737\,(12)       &   6.2152\,(50)       &   8.1596\,(17)       &   9.5068\,(33)       \\
  & sum   &    2.7629\,(12)       &  -0.1581\,(50)       &   1.7863\,(17)       &   3.1335\,(33)       \\[3pt]
5 & rad   &   -2.6027             &  -3.4394             &  -3.4394             &  -3.4394             \\
  & nrad  &    3.9135\,(63)       &   3.2304\,(64)       &   4.1985\,(23)       &   4.8699\,(40)       \\
  &       &    1.3109\,(63)       &  -0.2090\,(64)       &   0.7591\,(23)       &   1.4306\,(40)       \\[3pt]
6 & rad   &   -1.5340             &  -2.0456             &  -2.0456             &  -2.0456             \\
  & nrad  &    2.2606\,(43)       &   1.8855\,(80)       &   2.4381\,(50)       &   2.8206\,(43)       \\
  & sum   &    0.7265\,(43)       &  -0.1601\,(80)       &   0.3925\,(50)       &   0.7751\,(43)       \\
%
%
%
%
%
\end{tabular}
\end{ruledtabular}
\end{table*}

\begin{table*}
\caption{Coefficients of the asymptotic $1/n$ expansion (\ref{eq:ass}) for the
radiative (rad) and non-radiative (nrad) $m\alpha^6$ corrections of $1snd$
states of $^4$He, in units of $m\alpha^6$.
\label{tab:ass}}
\begin{ruledtabular}
\begin{tabular}{ldddddddd}
Coefficient &  \multicolumn{2}{c}{$n\,{}^1\!D$}
                          & \multicolumn{2}{c}{$n\,{}^3\!D_1$}
                                                   & \multicolumn{2}{c}{$n\,{}^3\!D_2$} &
                                                                    \multicolumn{2}{c}{$n\,{}^3\!D_3$} \\
 & \multicolumn{1}{c}{\ \ \ \ \ rad}  &  \multicolumn{1}{c}{\ \ \ \ \ nrad}
 & \multicolumn{1}{c}{\ \ \ \ \ rad}  &  \multicolumn{1}{c}{\ \ \ \ \ nrad}
 & \multicolumn{1}{c}{\ \ \ \ \ rad}  &  \multicolumn{1}{c}{\ \ \ \ \ nrad}
 & \multicolumn{1}{c}{\ \ \ \ \ rad}  &  \multicolumn{1}{c}{\ \ \ \ \ nrad}\\
\hline\\[-5pt]
 $c_0$ &    -0.3457  &  0.4874  & -0.4684  &  0.4098  &  -0.4684  &  0.5320    &  -0.4684  &  0.6093 \\
 $c_1$ &     0.0071  & -0.0144  & -0.0062  &  0.0427  &  -0.0062  & -0.0233    &  -0.0062  & -0.0002 \\
 $c_2$ &     0.4745  &  0.1159  &  0.9930  & -0.3627  &   0.9930  & -0.0637    &   0.9930  & -0.0134 \\
\end{tabular}
\end{ruledtabular}
\end{table*}

\begin{table*}
\caption{Individual contributions to theoretical ionization energies of the $1snd$ states of $^4$He, in MHz.
Fundamental constants are \cite{NIST} $R_{\infty}c = 3\,289\,841\,960.250\,8\,(64)$~MHz,
$\alpha^{-1} = 137.035\,999\,084\,(21)$, $ M/m = 7\,294.299\,541\,36$. Theoretical energies of
Morton {\em et al.}~\cite{morton:06:cjp} are rescaled for the updated value of the Rydberg constant.
\label{tab:energy}}
\begin{ruledtabular}
\begin{tabular}{ldddd}
                     &        \multicolumn{1}{c}{$4^1\!D_2$}
                                            &  \multicolumn{1}{c}{$4{}^3\!D_1$}
                                                                   &  \multicolumn{1}{c}{$4{}^3\!D_2$}
                                                                                          &  \multicolumn{1}{c}{$4^3\!D_3$}
                                                                   \\
                                                                   \hline\\[-5pt]
$E^{(2,0)}$            & -205811500.938       & -205870726.799       & -205870726.799       & -205870726.799       \\
$E^{(2,1)}$            &      28098.869       &      28250.067       &      28250.067       &      28250.067       \\
$E^{(2,2+)}$           &         -7.810       &         -7.677       &         -7.677       &         -7.677       \\
$E^{(4,0)}$            &       -527.343       &        -55.787       &       -602.728       &       -647.005       \\
$E^{(4,1)}$            &          0.108       &         -0.179       &          0.053       &          0.083       \\
$E^{(5,0)}$            &         -6.289       &         -7.426       &         -8.266       &         -7.722       \\
$E^{(5,1)}$            &         -0.002       &          0.002       &          0.003       &          0.002       \\
$E_{\rm MIX}$          &          7.684       &          0.          &         -7.684       &          0.          \\
$E^{(6,0)}_{\rm rad}$  &         -0.092       &         -0.119       &         -0.119       &         -0.119       \\
$E^{(6,0)}_{\rm nrad}$ &          0.143       &          0.116       &          0.152       &          0.177       \\
$E^{(7,0)}$            &          0.009\,(9)  &          0.011\,(11) &          0.011\,(11) &          0.011\,(11) \\
$E_{\rm FNS}$          &         -0.004       &         -0.005       &         -0.005       &         -0.005       \\
Total                  & -205783935.664\,(9)  & -205842547.795\,(11) & -205843102.990\,(11) & -205843138.986\,(11) \\
Morton 2006            & -205783935.809\,(9)  & -205842547.913\,(10) & -205843103.143\,(10) & -205843139.163\,(10) \\
Difference             &          0.145\,(12) &          0.119\,(15) &          0.153\,(15) &          0.178\,(15) \\
                                                                   \hline\\[-5pt]
                     &        \multicolumn{1}{c}{$5^1\!D_2$}
                                            &  \multicolumn{1}{c}{$5{}^3\!D_1$}
                                                                   &  \multicolumn{1}{c}{$5{}^3\!D_2$}
                                                                                          &  \multicolumn{1}{c}{$5^3\!D_3$}
                                                                   \\
                                                                   \hline\\[-5pt]
$E^{(2,0)}$            & -131697875.337       & -131732032.364       & -131732032.364       & -131732032.364       \\
$E^{(2,1)}$            &      17990.063       &      18077.238       &      18077.238       &      18077.238       \\
$E^{(2,2+)}$           &         -4.992       &         -4.916       &         -4.916       &         -4.916       \\
$E^{(4,0)}$            &       -321.862       &        -79.757       &       -359.518       &       -382.521       \\
$E^{(4,1)}$            &          0.068       &         -0.085       &          0.034       &          0.050       \\
$E^{(5,0)}$            &         -3.314       &         -3.989       &         -4.418       &         -4.140       \\
$E^{(5,1)}$            &         -0.001       &          0.001       &          0.001       &          0.001       \\
$E_{\rm MIX}$          &          3.490       &          0.          &         -3.490       &          0.          \\
$E^{(6,0)}_{\rm rad}$  &         -0.049       &         -0.064       &         -0.064       &         -0.064       \\
$E^{(6,0)}_{\rm nrad}$ &          0.073       &          0.060       &          0.078       &          0.091       \\
$E^{(7,0)}$            &          0.005\,(5)  &          0.006\,(6)  &          0.006\,(6)  &          0.006\,(6)  \\
$E_{\rm FNS}$          &         -0.002       &         -0.002       &         -0.002       &         -0.002       \\
Total                  & -131680211.860\,(5)  & -131714043.872\,(6)  & -131714327.415\,(6)  & -131714346.623\,(6)  \\
Morton 2006            & -131680211.934\,(5)  & -131714043.934\,(6)  & -131714327.494\,(6)  & -131714346.715\,(6)  \\
Difference             &          0.074\,(7)  &          0.061\,(8)  &          0.079\,(8)  &          0.092\,(8)  \\
                                                                   \hline\\[-5pt]
                     &        \multicolumn{1}{c}{$6^1\!D_2$}
                                            &  \multicolumn{1}{c}{$6{}^3\!D_1$}
                                                                   &  \multicolumn{1}{c}{$6{}^3\!D_2$}
                                                                                          &  \multicolumn{1}{c}{$6^3\!D_3$}
                                                                   \\
                                                                   \hline\\[-5pt]
$E^{(2,0)}$            &  -91445943.507       &  -91466919.734       &  -91466919.734       &  -91466919.734       \\
$E^{(2,1)}$            &      12497.472       &      12551.001       &      12551.001       &      12551.001       \\
$E^{(2,2+)}$           &         -3.463       &         -3.416       &         -3.416       &         -3.416       \\
$E^{(4,0)}$            &       -206.310       &        -66.005       &       -227.805       &       -241.222       \\
$E^{(4,1)}$            &          0.044\,(1)  &         -0.046       &          0.023       &          0.032       \\
$E^{(5,0)}$            &         -1.947       &         -2.366       &         -2.614       &         -2.454       \\
$E^{(5,1)}$            &         -0.001       &          0.001       &          0.001       &          0.001       \\
$E_{\rm MIX}$          &          1.902       &          0.          &         -1.902       &          0.          \\
$E^{(6,0)}_{\rm rad}$  &         -0.029       &         -0.038       &         -0.038       &         -0.038       \\
$E^{(6,0)}_{\rm nrad}$ &          0.042       &          0.035       &          0.045       &          0.053       \\
$E^{(7,0)}$            &          0.003\,(3)  &          0.004\,(4)  &          0.004\,(4)  &          0.004\,(4)  \\
$E_{\rm FNS}$          &         -0.001       &         -0.001       &         -0.001       &         -0.001       \\
Total                  &  -91433655.795\,(3)  &  -91454440.567\,(4)  &  -91454604.437\,(4)  &  -91454615.775\,(4)  \\
Morton 2006            &  -91433655.838\,(3)  &  -91454440.602\,(4)  &  -91454604.483\,(4)  &  -91454615.829\,(4)  \\
Difference             &          0.043\,(4)  &          0.035\,(5)  &          0.046\,(5)  &          0.054\,(5)  \\
\end{tabular}
\end{ruledtabular}
\end{table*}

\begin{table*}
\caption{Theoretical ionization energies of the $1snd$ states of $^4$He, in MHz.
\label{tab:energylist}}
\begin{ruledtabular}
\begin{tabular}{ldddd}

$n$              &        \multicolumn{1}{c}{n$^1\!D_2$}
                                            &  \multicolumn{1}{c}{$n{}^3\!D_1$}
                                                                   &  \multicolumn{1}{c}{$n{}^3\!D_2$}
                                                                                          &  \multicolumn{1}{c}{$n^3\!D_3$}
                                                                   \\
                                                                   \hline\\[-5pt]
  3  &    -365\,917\,748.661\,(19)  &-366\,018\,892.691\,(23)  &-366\,020\,217.716\,(23)  &-366\,020\,292.981\,(23) \\
  4  &    -205\,783\,935.6645\,(86) &-205\,842\,547.795\,(11)  &-205\,843\,102.990\,(11)  &-205\,843\,138.986\,(11) \\
  5  &    -131\,680\,211.8599\,(46) &-131\,714\,043.8724\,(60) &-131\,714\,327.4151\,(60) &-131\,714\,346.6227\,(60) \\
  6  &     -91\,433\,655.7952\,(30) & -91\,454\,440.5666\,(36) & -91\,454\,604.4375\,(36) & -91\,454\,615.7752\,(36) \\
  7  &     -67\,169\,717.1272\,(20) & -67\,183\,264.5655\,(20) & -67\,183\,367.6781\,(20) & -67\,183\,374.8986\,(20) \\
  8  &     -51\,423\,248.1215\,(10) & -51\,432\,523.2304\,(20) & -51\,432\,592.2710\,(20) & -51\,432\,597.1421\,(20) \\
  9  &     -40\,628\,480.2532\,(9)  & -40\,635\,090.4352\,(10) & -40\,635\,138.9071\,(10) & -40\,635\,142.3441\,(10) \\
 10  &     -32\,907\,601.9048\,(6)  & -32\,912\,470.7471\,(9)  & -32\,912\,506.0729\,(9)  & -32\,912\,508.5867\,(9) \\
\end{tabular}
\end{ruledtabular}
\end{table*}

\begin{table*}
  \caption{Comparison of different theoretical predictions with experimental results for transition energies
  involving $nD$ states
  in $^4$He, in MHz. The experimental value for the $3\,^1\!D$ -- $3\,^3\!D_1$ transition energy
  was obtained by combining several measurements for different transitions
  \cite{huang:18,rengelink:18,zheng:17,luo:16}.
   \label{tab:comp}
}
\begin{center}
\begin{ruledtabular}
\begin{tabular}{l w{11.6} c w{11.6} w{2.6} w{11.6} w{2.6}}
\centt{Transition}
                            & \centt{Experiment}
                            & \centt{Ref.}
                            & \centt{Present theory}
                            & \centt{Difference}
                            & \centt{Morton 2006}
                            & \centt{Difference}
                            \\
                            &
                            &
                            &
                            &  \centt{from experiment}
                            & \centt{\cite{morton:06:cjp}}
                            &  \centt{from experiment}
                            \\\hline
\\[-1.0ex]
  $3\,^1\!D$ -- $2\,^1\!S$            & 594\,414\,291.803\,(13)
                            &\cite{huang:18}
                            & 594\,414\,289.3\,(1.9)
                            &             2.5\,(1.9)
                            & 594\,414\,292.\,(5.)
                            &             0.\,(5.)
                            \\ [0.5ex]
  $3\,^3\!D_1$ -- $2\,^3\!S$          & 786\,823\,850.002\,(56) \,
                            &\cite{dorrer:97}
                            & 786\,823\,848.7\,(1.3)\,
                            &             1.3\,(1.3)
                            & 786\,823\,845.\,(7.)
                            &             4.\,(7.)
                            \\ [0.5ex]
  $3\,^3\!D_1$ -- $2\,^3\!P_0$        & 510\,059\,755.352\,(28) \,
                            &\cite{luo:16}
                            & 510\,059\,754.2\,(0.7)\,
                            &             1.2\,(0.7)\,
                            & 510\,059\,749.\,(2.)
                            &             6.\,(2.)
                            \\ [0.5ex]
  $3\,^1\!D$ -- $2\,^1\!P$            & 448\,791\,399.11\,(27)\,
                            &\cite{luo:13:b}
                            & 448\,791\,397.8\,(0.4)\,
                            &             1.3\,(0.5)
                            & 448\,791\,400.5\,(2)
                            &            -1.4\,(2)
                            \\  [0.5ex]
 $7\,^1\!D$ -- $2\,^1\!S$             & 893\,162\,323.88\,(12) & \cite{lichten:91}
                            & 893\,162\,320.9\,(1.9)
                            &             3.0\,(1.9)
                            & 893\,162\,324.\,(5.)
                            &             0.\,(5.)
                            \\[0.5ex]
 $8\,^1\!D$ -- $2\,^1\!S$             & 908\,908\,792.76\,(7)  & \cite{lichten:91}
                            & 908\,908\,789.9\,(1.9)
                            &             2.9\,(1.9)
                            & 908\,908\,793.\,(5.)
                            &             0.\,(5.)
                            \\[0.5ex]
 $9\,^1\!D$ -- $2\,^1\!S$             & 919\,703\,560.56\,(11) & \cite{lichten:91}
                            & 919\,703\,557.7\,(1.9)
                            &             2.9\,(1.9)
                            & 919\,703\,561.\,(5.)
                            &             0.\,(5.)
                            \\[0.5ex]
 $10\,^1\!D$ -- $2\,^1\!S$            & 927\,424\,439.05\,(10) & \cite{lichten:91}
                            & 927\,424\,436.1\,(1.9)
                            &             3.0\,(1.9)
                            & 927\,424\,439.\,(5.)
                            &             0.\,(5.)
                            \\[0.5ex]
  $5\,^3\!D_2$ -- $2\,^3\!P_2$
                            & 744\,396\,218.(7.) & \cite{ross:20}
                            & 744\,396\,227.7\,(0.7)
                            &            10.(7.)
                            & 744\,396\,231.(2.)
                            &            13.(7.)
                            \\[0.5ex]
  $3\,^1\!D$ -- $3\,^3\!D_1$          & 101\,143.943\,(31) \,
                            & \cite{huang:18,rengelink:18,zheng:17,luo:16}
                            & 101\,144.029\,(23)
                            &        0.086\,(37)
                            & 101\,143.95\,(3)
                            &        0.01\,(4)
                            \\[0.5ex]
  $5\,^1\!D$ -- $5\,^3\!D_1$
                            & 33\,850.(10.) & \cite{ross:20}
                            & 33\,832.013\,(8)
                            &      18.(10.)
                            & 33\,832.000\,(8)
                            &      18.(10.)
                            \\[0.5ex]
  $5\,^3\!D_1$ -- $5\,^3\!D_2$
                            & 279.(10.) & \cite{ross:20}
                            & 283.543\,(8)
                            &   5.(10.)
                            & 283.560\,(8)
                            &   5.(10.)
                            \\[0.5ex]
  $5\,^3\!D_1$ -- $5\,^3\!D_3$
                            & 297.(10.) & \cite{ross:20}
                            & 302.750\,(8)
                            &   6.(10.)
                            & 302.781\,(8)
                            &   6.(10.)
                            \\[0.5ex]
\end{tabular}
\end{ruledtabular}
\end{center}
\end{table*}
%


\end{document}